\documentclass[epj,nopacs]{modsvjour}
\usepackage{graphics}
\begin{document}
\title{Final State Correlations at LEP 2}
\subtitle{Bose-Einstein Correlations and the W Mass}
\author{Paul de Jong% etc
% \thanks is optional - remove next line if not needed
\thanks{e-mail: paul.de.jong@nikhef.nl}
% \thanks{\emph{Present address:} Insert the address here if needed}%
}                     % Do not remove
%
% \offprints{}          % Insert a name or remove this line
%
\institute{NIKHEF, P.O. Box 41882, 1009 DB Amsterdam, the Netherlands}
\date{September 3, 2003. Presented at EPS HEP2003, Aachen, July 2003.}
% The correct dates will be entered by Springer
%
\abstract{
% Insert your abstract here.
Recent experimental results on Bose-Einstein correlations are presented. 
Emphasis will be
put on the measurement of between-W correlations in WW events at LEP 2.
%
%\PACS{
%      {PACS-key}{discribing text of that key}   \and
%      {PACS-key}{discribing text of that key}
%     } % end of PACS codes
} %end of abstract
\maketitle
\section{Introduction}
\label{pdjsec:intro}
The W mass and width measurements at LEP 2 rely on good Monte Carlo
simulations of physics and detectors in the
$e^+ e^- \rightarrow$ WW process. 
The Monte Carlo model parameters
are tuned to data, and the detector simulations are calibrated against
large samples of Z events at LEP 1 and the calibration periods of
LEP 2. 
Remaining sources of uncertainty are, among others, the
simulation of fragmentation and hadronization, in particular colour
interactions and correlations between partons and particles from
the decay of different W's in WW $\rightarrow qqqq$ events.
If the Monte Carlo does not simulate this correctly, a bias in this
channel may result. 
The four LEP experiments try to estimate these
interconnections from the data, and this article will describe recent
studies of Bose-Einstein correlations (BEC).

\section{Bose-Einstein Correlations}
\label{pdjsec:bec}
The observed enhancement of the production of identical bosons close
in phase space is considered to be a result of the requirement of
symmetrization of the production amplitude. We define a two-particle
density function $\rho_2 (Q)$ as $\rho_2 (Q) = 1/N_{\mathrm{ev}} 
dn_{\mathrm{pairs}}/dQ$, where $Q = \sqrt{-(p_1 - p_2)^2} =
\sqrt{M^2 - 4m^2}$ for pairs of identical bosons with 4-momenta
$p_1$ and $p_2$ and mass $m$. The correlation function $R(Q)$ is
then defined as $R(Q) = \rho_2(Q) / \rho_2^{\mathrm{ref}}(Q)$, where
$\rho_2^{\mathrm{ref}}(Q)$ is derived from a reference sample with
all properties of the sample under study, except Bose-Einstein
correlations. 
Such a sample is difficult to obtain; 
analyses have typically used reference samples of mixed events or 
unlike-sign particles, each
of these have their disadvantages. It is known that for a spherical
and Gaussian source with radius $r$, $R(Q)$ can be written as 
\begin{equation}
\label{pdjeq1}
R(Q) = N (1 + \delta Q) (1 + \lambda \exp (-(rQ)^2)),
\end{equation}
where $N$ is
a normalization, $\delta$ describes long-range (non-BEC) correlations,
and $\lambda$ is the correlation strength (or `coherence' or
`chaoticity' parameter).

BEC between particles from the decay of a single W (inside-W-BEC) are
identical to BEC in Z events, if corrected for the flavour difference.
Studies of Z events and deep-inelastic scattering data have found that:
\begin{itemize}
\item correlations between more than two particles exist~\cite{cor3};
\item $\pi^0 \pi^0$ correlations exist, even though some 97\% of the $\pi^0$'s
in these correlations originate from the decay of different 
hadrons~\cite{opalpi0};
\item generalized BEC may exist in $\pi^{\pm} \pi^0$ or $\pi^+ \pi^-$ 
pairs~\cite{genbec};
\item the source is not spherical, but elongated~\cite{bec2d}.
\end{itemize}

% \subsection{Subsection title}
% \label{sec:2}
\section{Monte Carlo Implementation}
\label{pdjsec:mc}
The implementation of BEC in Monte Carlo's can be categorized in
three classes:
\begin{itemize}
\item PYTHIA (LUBOEI)~\cite{luboei};
\item global reweighting methods~\cite{reweight};
\item Lund string fragmentation inspired models~\cite{lund}.
\end{itemize}
At the time of analysis, only PYTHIA was available as a mature MC, and
experiments compare their data to PYTHIA, with
either only inside-W BEC, or both inside-W and between-W BEC (``full'' PYTHIA).
The PYTHIA parameters corresponding to $\lambda$ and $r$ are obtained
by tuning the Monte Carlo to Z events (without Z $\rightarrow b \bar{b}$), 
and are also suited for
inside-W BEC. In the analyses presented here, the parameters for
between-W BEC simulations have been taken to be the same as for inside-W BEC.
Variant BE$_{32}$ is used. 

Recently, the ALFS Monte Carlo has appeared as an implementation
of BEC in the Lund model~\cite{lund}.

\section{Between-W Correlations Measurement}
\label{pdjsec:betweenw}

\subsection{Method}
\label{pdjsubsec:method}
The method uses a reference sample consisting of mixed semi-hadronic
WW events (WW $\rightarrow qq \ell \nu$). Mixed events are
constructed by taking two semi-hadronic WW events from a pool, removing
the W-decay leptons from the events, and rotating and boosting the W's
such that they are back-to-back. Care has to be taken in subtracting
the non-WW four-jet background from the $qqqq$ sample.
Mixed events have by construction no
between-W BEC, and have the same inside-W BEC as real $qqqq$ events. The
between-W BEC can now be extracted by comparing the
real $qqqq$ to the mixed events~\cite{method}:
\begin{displaymath}
\Delta \rho (Q) = \rho_2^{WW}(Q) - 2 \rho_2^W(Q) - 
2 \rho_{\mathrm{mix}}^{WW}(Q),
\end{displaymath}
\begin{displaymath}
D(Q) = \rho_2^{WW}(Q) / (2 \rho_2^W(Q) + 2 \rho_{\mathrm{mix}}^{WW}(Q)),
\end{displaymath}
where $\rho_2^{WW}$ is the two-particle density function in $qqqq$ events,
$\rho_2^W$ is that function within single W's  taken from $qq \ell \nu$
events, and $\rho_{\mathrm{mix}}^{WW}$ is that function for pairs of
particles from different W's in mixed events.
We also define $\Delta \rho'$ and $D'$ as $\Delta \rho$ and $D$ from data
minus (c.q. divided by) PYTHIA without between-W BEC. 
If between-W BEC are absent, $\Delta \rho = 0$ and $D = 1$. 
Experiments apply a phenomenological fit to the $D(Q)$ distribution 
similar to Equation~\ref{pdjeq1} (or like $\lambda \exp (-r Q)$) in order
to quantify the between-W BEC strength. However, $D(Q)$ is not a 
correlation function like $R(Q)$, and parameters should be interpreted 
with care.

As an alternative, between-W BEC measurements are quantified
by integration of the $\Delta \rho(Q)$ distribution, see the experimental
papers for results.

\subsection{L3 Results}
\label{pdjsubsec:l3}

\begin{figure}
\begin{center}
\resizebox{0.49\textwidth}{!}{%
  \includegraphics{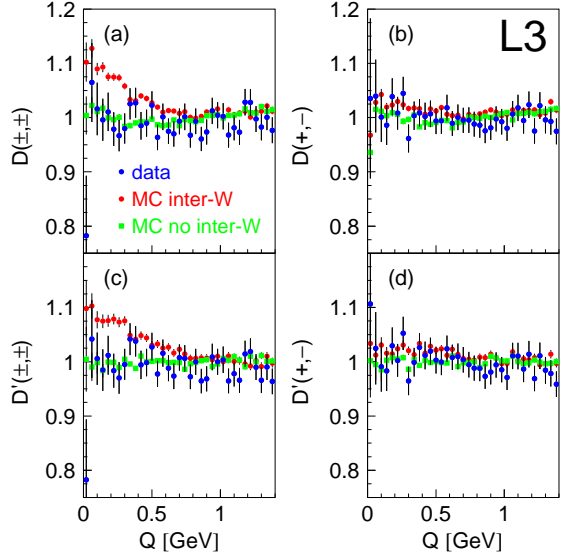}
}
\end{center}
\caption{The L3 $D(Q)$ and $D'(Q)$ distributions for like-sign 
and unlike-sign pairs in data and in PYTHIA.}
\label{pdjfig:l3fig3}       % Give a unique label
\end{figure}

The final L3 results~\cite{l3wwbec} use 629 pb$^{-1}$ of data between 
$\sqrt{s} =$ 189 and 209 GeV, giving some 3800 $qq \ell \nu$ and 
5100 $qqqq$ events.
The $D(Q)$ and $D'(Q)$ distributions are shown in Figure~\ref{pdjfig:l3fig3}.

The L3 results are consistent with no between-W BEC, and disagree with
full PYTHIA at the 3.8 $\sigma$ level.

\subsection{DELPHI Results}
\label{pdjsubsec:delphi}

\begin{figure}
\begin{center}
\resizebox{0.40\textwidth}{!}{%
  \includegraphics{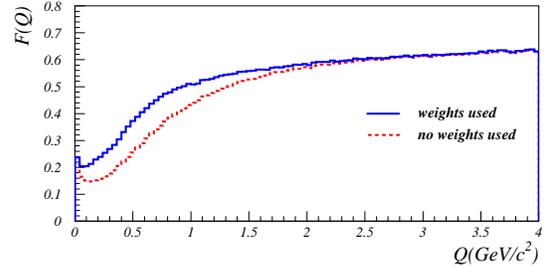}
}
\end{center}
\caption{The fraction $F(Q)$ of pion pairs where the two pions
originate from different W's, as a
function of $Q$, before and after reweighting (DELPHI).}
\label{pdjfig:delphifig7}       % Give a unique label
\end{figure}

DELPHI~\cite{delphiwwbec} observe that at low $Q$, the fraction 
$F(Q)$ of pion pairs where the two pions originate
from different W's is very low, as shown in Figure~\ref{pdjfig:delphifig7}.
In order to increase the sensitivity of the between-W correlations
measurement, DELPHI reweight the pairs with their information content,
obtained from three variables sensitive to the W parent.

\begin{figure}
\begin{center}
\resizebox{0.40\textwidth}{!}{%
  \includegraphics{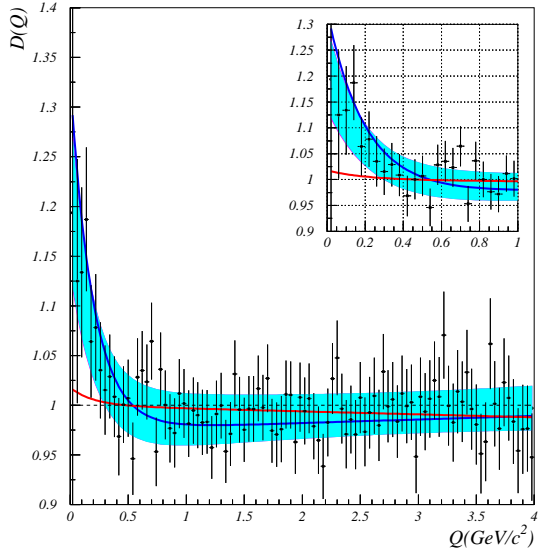}
}
\end{center}
\caption{The DELPHI $D(Q)$ distribution for like-sign pairs, and the 
fit to this distribution (band), compared
to full PYTHIA (thick curve) and PYTHIA with inside-W BEC only.}
\label{pdjfig:delphifig8}       % Give a unique label
\end{figure}

For the analysis, DELPHI use 550 pb$^{-1}$ of data between $\sqrt{s} =$ 189
and 209 GeV, giving 2567 $qq \ell \nu$ and 3252 $qqqq$ events.
The $D(Q)$ distribution is shown in Figure~\ref{pdjfig:delphifig8}.

DELPHI observe an indication for between-W BEC
with a significance corresponding to 2.9 standard deviations. The
magnitude of the effect is 2/3 of full PYTHIA. DELPHI
also observe this in the unlike-sign pairs. The between-W BEC effect
appears to be situated at smaller $Q$, or larger $r$, than in full PYTHIA. 

\subsection{ALEPH Results}
\label{pdjsubsec:aleph}

\begin{figure}
\begin{center}
\resizebox{0.49\textwidth}{!}{%
  \includegraphics{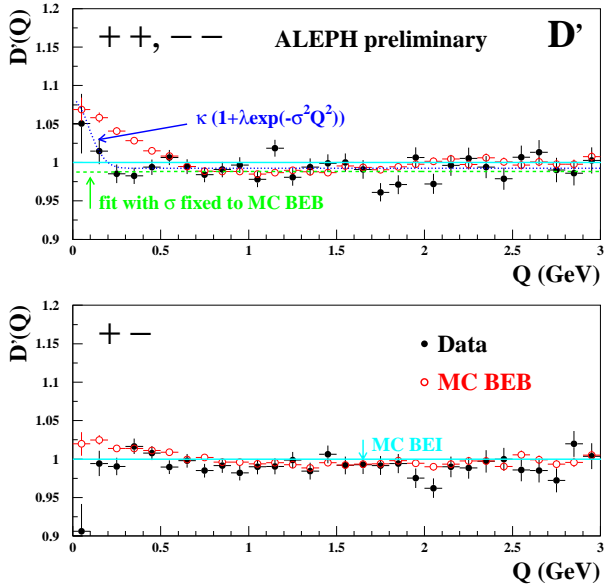}
}
\end{center}
\caption{The ALEPH $D'(Q)$ distributions for like-sign and unlike-sign
pairs and the fits with $\sigma = r$ left
free or fixed to the full PYTHIA (= MC BEB) value.}
\label{pdjfig:alephfig2}       % Give a unique label
\end{figure}

ALEPH~\cite{alephwwbec} use 685 pb$^{-1}$ of data between $\sqrt{s} =$ 183
and 209 GeV, giving 6154 $qqqq$ events, and 2406 constructed mixed events.
The $D'(Q)$ distributions are shown in Figure~\ref{pdjfig:alephfig2}.

ALEPH observe no between-W BEC in the $\Delta \rho'$ and $D'$ distributions
if $r$ is fixed to the full PYTHIA value,
and disagree with full PYTHIA at the 3.7 standard deviation level.
If $r$ is left free in the fit to $D'(Q)$, a preference for larger $r$ 
in between-W BEC is seen than in full PYTHIA.

\subsection{OPAL Results}
\label{pdjsubsec:opal}

\begin{figure}
\begin{center}
\resizebox{0.49\textwidth}{!}{%
  \includegraphics{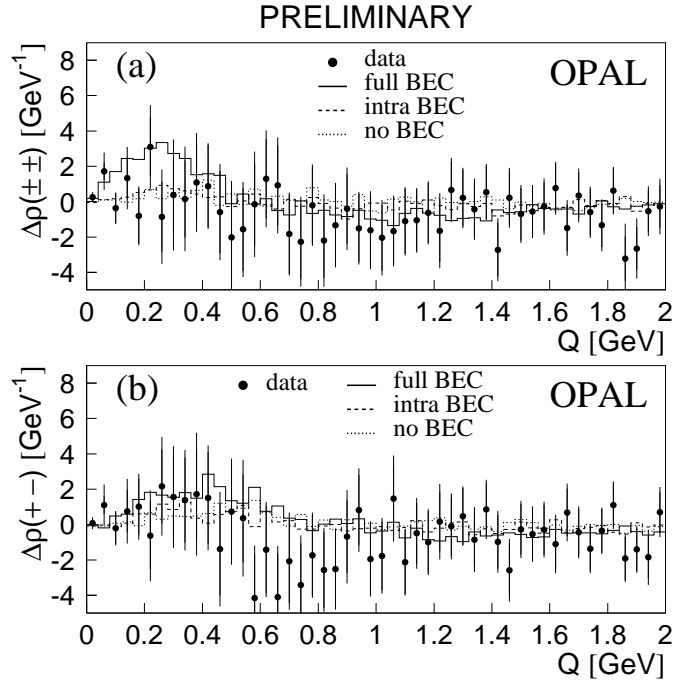}
}
\end{center}
\caption{The OPAL $\Delta \rho (Q)$ distributions for like-sign and unlike-sign
pairs.}
\label{pdjfig:opalfig6ab}       % Give a unique label
\end{figure}

OPAL~\cite{opalwwbec} use 680 pb$^{-1}$ of data between $\sqrt{s} =$ 183
and 209 GeV, giving 4533 $qq \ell \nu$ and 4470 $qqqq$ events.
The $\Delta \rho (Q)$ distributions are shown in Figure~\ref{pdjfig:opalfig6ab}.

OPAL compare their data to both PYTHIA scenario's, and find that
the results for $\Delta \rho$ prefer no between-W BEC, whereas
the $D$ analysis is consistent with either scenario.

\subsection{LEP Combination}
\label{pdjsubsec:lepcombi}
\begin{figure}
\begin{center}
\resizebox{0.40\textwidth}{!}{%
  \includegraphics{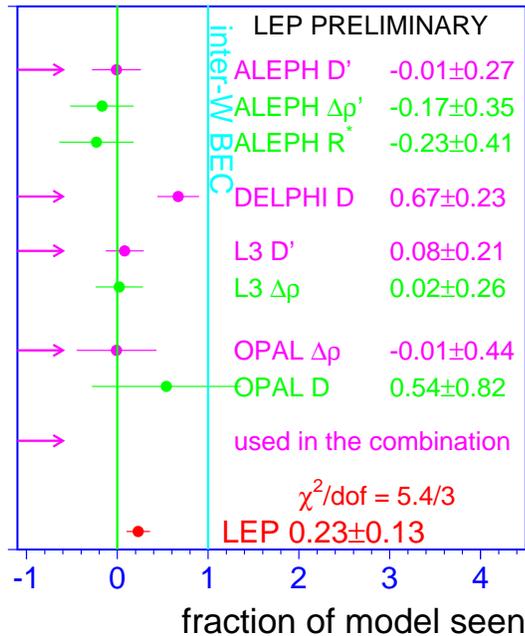}
}
\end{center}
\caption{LEP combination of the measured between-W BEC strengths expressed
as fraction of full PYTHIA.}
\label{pdjfig:lepcombi}       % Give a unique label
\end{figure}

Since the measurements are statistics-limited, it is interesting to
combine them. The combination is shown in Figure~\ref{pdjfig:lepcombi},
where the measured between-W BEC strengths in each experiment are 
expressed as fraction of full PYTHIA.
The arrows mark the results used in the combination. The combination
has a $\chi^2$ of 5.4 for 3 degrees of freedom; the probability for
such a $\chi^2$ (or higher) is 15\%, which is acceptable. The largest
deviation from the average (DELPHI) is less than two sigma.

The combination indicates that the LEP experiments measure a between-W
correlation strength of $0.23 \pm 0.13$ times the one of full 
PYTHIA~\footnote{If the OPAL $D(Q)$ result had been used in the combination
instead of the $\Delta \rho(Q)$ result, this number would have been only
marginally different: $0.25 \pm 0.14$}.
This would correspond, again in the PYTHIA framework, to a W mass shift
in the $qqqq$ channel
of 8 $\pm$ 5 MeV, and a W width shift of some 12 $\pm$ 8 MeV.
The observation that $r$ for between-W BEC seems larger than in full
PYTHIA is interesting. Its effects remain to be further studied, but they
again point to a W mass shift smaller than predicted by full PYTHIA.

\section{Conclusions}
\label{pdjsec:conclusions}
For the first time, the four LEP experiments have used the same method 
to measure between-W BEC in WW data at LEP 2; the DELPHI, ALEPH and OPAL
results are still preliminary.
The experiments measure in data a between-W BEC strength of $0.23 \pm 0.13$
times the implementation of full PYTHIA, which in the framework of that model
corresponds to an upper limit on the W mass shift in the
$qqqq$ channel of 13 MeV at 68\% CL.
Other MC models also predict shifts between 0 and 
15 MeV~\cite{reweight,lund}.
There are indications that the small between-W BEC effect is located at 
smaller $Q$, or larger $r$,
than BEC in single W or Z events.
The data are consistent with the emerging theoretical picture that between-W
BEC from incoherent W decays probably exist, but that the effects are much
suppressed w.r.t. inside-W BEC (and thus full PYTHIA): there are two
separated sources, and at low $Q$ few pairs of pions originate 
from different W's.
It thus appears that the influence of between-W BEC on the W mass is
limited.
% For two-column wide figures use
%%\begin{figure*}
% Use the relevant command for your figure-insertion program
% to insert the figure file. See example above.
% If not, use
%%\vspace*{5cm}       % Give the correct figure height in cm
%%\caption{Please write your figure caption here}
%%\label{fig:2}       % Give a unique label
%%\end{figure*}
%
% For tables use
% \begin{table}
% \caption{Please write your table caption here}
% \label{tab:1}       % Give a unique label
% For LaTeX tables use
% \begin{tabular}{lll}
% \hline\noalign{\smallskip}
% first & second & third  \\
% \noalign{\smallskip}\hline\noalign{\smallskip}
% number & number & number \\
% number & number & number \\
% \noalign{\smallskip}\hline
% \end{tabular}
% Or use
% \vspace*{5cm}  % with the correct table height
% \end{table}
%
% BibTeX users please use
% \bibliographystyle{}
% \bibliography{}
%
% Non-BibTeX users please use

\end{document}